
\input phyzzx

%
%
\newcount\lemnumber   \lemnumber=0
\newcount\thnumber   \thnumber=0
\newcount\conumber   \conumber=0

\def\myeq{{\rm \chapterlabel\the\equanumber}}

\def\Lemma{\par\noindent\global\advance\lemnumber by 1
           {\bf Lemma\ (\chapterlabel\the\lemnumber)}}
\def\Corollary{\par\noindent\global\advance\conumber by 1
           {\bf Corollary\ (\chapterlabel\the\conumber)}}
\def\Theorem{\par\noindent\global\advance\thnumber by 1
           {\bf Theorem\ (\chapterlabel\the\thnumber)}}

%
%
\def\e{\adveq\eqno{\rm (\chapterlabel\the\equanumber)}}
\def\adveq{\global\advance\equanumber by 1}
\def\twoline#1#2{\displaylines{\qquad#1\hfill(\adveq\myeq)\cr\hfill#2
\qquad\cr}}

%
%
\font\tensl=cmsl10
\font\tenss=cmssq8 scaled\magstep1
\outer\def\quote{
   \begingroup\bigskip\vfill
   \def\endquote{\endgroup\eject}
    \def\par{\ifhmode\/\endgraf\fi}\obeylines
    \tenrm \let\tt=\twelvett
    \baselineskip=10pt \interlinepenalty=1000
    \leftskip=0pt plus 60pc minus \parindent \parfillskip=0pt
     \let\rm=\tenss \let\sl=\tensl \everypar{\sl}}
\def\from#1(#2){\smallskip\noindent\rm--- #1\unskip\enspace(#2)\bigskip}

\def\WIS{\address{Department of Physics\break
      Weizmann Institute of Science\break
      Rehovot 76100, Israel}}

\def\r#1{$\lb \rm#1 \rb$}

%
%
\def\rarrow{\rightarrow}
\def\d#1{{\rm d}#1\,}

\def\ch{{\rm ch}}

\def\semidirect{\mathrel{\raise0.04cm\hbox{${\scriptscriptstyle |\!}$
\hskip-0.175cm}\times}}

\def\mod{\mathop{\rm mod}\nolimits}

\def\ref#1{$^{#1}$}

\def\pr#1{{#1^\prime}}

\def\half{{1\over2}}
\def\lb{\lbrack}
\def\rb{\rbrack}

\def\pr{\prime}

\overfullrule=0pt
\def\pr#1{{#1}^\prime}\def\Mu{\Omega}
\Pubnum={WIS-92/34/March-Ph}
\date{March, 1992}
\titlepage
\title{Symplectic Fusion Rings and their Metric}
\author{Doron Gepner and Adam Schwimmer\foot{Additional address for
Adam Schwimmer: SISSA and INFN, Trieste, Italy.}}
\WIS
\abstract
The fusion of fields in a rational conformal field theory gives rise
to a ring structure which has a very particular form. All such rings
studied so far were shown to arise from some potentials. In this paper
the fusion rings of the WZW models based on the symplectic group are
studied. It is shown that they indeed arise from potentials which are
described. These potentials give rise to new massive perturbations of
superconformal hermitian symmetric models. The metric of the perturbation
is computed and is shown to be given by solutions of the sinh--gordon
equation. The kink structure of the theories is described, and it is argued
that these field theories are integrable. The $S$ matrices for the fusion
theories are argued to be non--minimal extensions of the $G_k\times G_1/
G_{k+1}$ $S$ matrices with the adjoint perturbation, in the case of
$G=SU(N)$.
\endpage
%

One of the central objects in the study of rational conformal field theory
are the fusion rules which specify how the different fields in the theory
fuse in the operator product algebra. For the WZW models the fusion rules
were determined via a group theoretic criteria in ref.
\REF\GW{D. Gepner and E. Witten, Nucl. Phys. B278 (1986) 493.}\r\GW.
These assume the form
$$C^\lambda C^\mu = \sum_\nu N_{\lambda,\mu}^\nu C^\nu \e $$
where $N_{\lambda,\nu}^\mu$ are non--negative integers which express
the number of times that the block of fields $C^\nu$ appear in the
product. $\lambda,\nu,\mu$ denote highest weight vectors at level
$k$, $\nu \theta\leq k$ where $k\geq0$ is the integral coefficient of the
Wess--Zumino term. Under this product, eq. (1), the primary fields form a
commutative ring
with a unit. In addition, the vacuum expectation value in the theory gives
rise to an invariant bilinear form
$$\langle C^\lambda C^\mu \rangle =\delta_{\lambda,\bar\mu} \e $$
where $\bar\mu$ is the conjugate representation (charge conjugation).
A connection between the fusion rules and the matrix of modular
transformation was described in ref. \REF\Verlinde{E. Verlinde,
Nucl. Phys. B300 [FS22] 360 (1988).}\r\Verlinde.

A well known description for commutative rings is as a quotient of a free
polynomial algebra modulo some relations, $P\lb x_i \rb /(p_i)$ where
$p_i(x)$ denotes the relations among the generators $x_i$. Among such
rings are the Jacobian varieties where the relations obey $p_i=\partial V(x)
/\partial x_i$. The Jacobian rings are very special and the generic ring
is far from being one\foot{We thank D. Kazhdan for a discussion of this
point.}. In ref. \REF\FR{D. Gepner, Comm. Math. Phys. 141
(1991) 381.}\r\FR\ the structure of the fusion rings of
$SU(N)_k$ was described as a Schubert type calculus. It was shown that
these rings are indeed Jacobian varieties with a potential which is
$$V(x_r)=\sum_{i=1}^N q_i^{N+k}, {\rm where\ } x_r=\sum_{i_1<i_2<
\ldots <i_r} q_{i_1} q_{i_2} q_{i_3}\ldots q_{i_r}\e$$
and $x_N=1$. For example, in the case of $SU(2)$, $V$ becomes the
Chebyshev polynomial of the first kind, $T_{k+2}(x)$
where $T_n(2\cos\theta)=2\cos(n\theta)$.

The fact that the fusion ring is a Jacobian variety has a considerable
geometrical significance. It implies essentially that it describes the
moduli space of the complex manifold $V(x_i)=0$,
where $x_i$ are any complex numbers, $x_i\in C$.
Further, it allows one to describe
it as a chiral algebra of a massive $N=2$ supersymmetric field theory,
which are themselves geometrical in nature \REF\CV{S. Cecotti and C. Vafa,
Nucl. Phys. B367 (1991) 359.}\r{\FR,\CV}.
Basing on the known examples, it was conjectured in ref. \r\FR\ that all the
fusion rings of rational conformal field theory are Jacobian varieties.
If proven correct, this would allow one to classify all such conformal field
theories via their potentials.
In this note the fusion rings of the symplectic algebra $C_n$ are investigated.
It is shown that they are indeed Jacobian varieties for any $n$ and any $k$,
and the potential is described.

Let us recall theorem (4.1) in ref. \r\FR. According to it any fusion ring
is of the
form $P\lb x_\alpha \rb /I$ where $x_\alpha$ is a set of generating primary
fields and the ideal $I$ can be described as the set of polynomials vanishing
at the points
$$x_\alpha=S_{\alpha,\beta}^\dagger/S_{0,\beta},\e$$
where $\beta$ ranges over all the primary fields, and $S_{\alpha,\beta}$
is the matrix of modular transformations. Further, any ideal which
vanishes precisely at these points is identical to $I$.

Thus, in order to establish that $I=(\partial_i V)$ where $V$ is some
potential, it is enough to establish that the extrema of $V$ are
precisely given by eq. (4). In the case of a current algebra theory
based on the Lie algebra $G$, substituting the value of the matrix of
modular transformations,
$$S_{\lambda,\mu}={\left\vert{M^*\over (k+g)M}\right\vert}^\half
\sum_{w\in W} (-1)^w e^{-2\pi iw(\lambda+\rho)
(\mu+\rho)/(k+g)},\e$$
we find that eq. (4) assumes the form
$$x_{\lambda}={\sum_{w\in W} (-1)^w e^{2\pi iw(\lambda+\rho)(\mu+\rho)
/(k+g)} \over \sum_{w\in W} (-1)^w e^{2\pi i w(\rho)(\mu+\rho)/(k+g)} },\e $$
where $\lambda$ is one of the fundamental weights, $\mu$ is any weight at
level $k$, $\rho$ is half the sum of positive roots, and $W$ denotes the
Weyl group. Using the Weyl character formula, eq. (6) can be described as
the specialization of the finite group character,
$$x_{\lambda}=\ch_\lambda\left({\mu+\rho\over k+g}\right),\e$$
where
$$\ch_\lambda(\alpha)=\sum_{\mu\in L(\lambda)} e^{2\pi i\mu\alpha}=
{\sum_{w\in W} (-1)^w e^{2\pi iw(\lambda+\rho)\alpha}\over
 \sum_{w\in W} (-1)^w e^{2\pi iw(\rho)\alpha} }.\e$$
The problem of finding a fusion potential thus reduces to finding
a polynomial whose extrema give the specialization points
$(\mu+\rho)/(k+g)$. Note that this is equivalent to finding a
fusion potential for the bosonic theory on the lattice $\sqrt{k+g} M_l$,
where $M_l$ is the long root lattice, which is Weyl invariant
\REF\Cres{M. Crescimanno, MIT preprint, MIT-CTP-2021, October 1991.}\r\Cres.
Such a potential will give precisely these extrema points,
where the weights which are not strictly dominant do not contribute
since the Jacobian vanishes on them. Note however that it is crucial
that the potential will be Weyl invariant, otherwise it cannot be
expressed as a polynomial in the symmetric variables $x_\alpha$.

Let us consider then the bosonic models. Here we have $r$ free
bosons propagating on a torus which is specified by the lattice
$\sqrt{k} M$ where $M$ is an even lattice ($\alpha^2=2$ for all
$\alpha$ in $M$), which is
generated by the vectors $l_j$.
The primary fields are of the form $A_p=\exp(\vec p \vec \phi/\sqrt k)$ and are
specified by the set of the allowed momenta $\vec p$. Since this field
must be single valued on the torus, it follows that $\vec p\in M^*$,
where $M^*$ denotes the dual lattice, which is generated by $m_i$ where
$m_i l_j=\delta_{ij}$. Now, since the dimension of the field $A_p$ is
$p^2/(2k)$ the extended algebra of the model consists
of all the fields $A_p$ where $p\in M$. It follows that the primary
fields are given by $p\in M^* \mod k M$, or $A_p=A_{p+m}$ for any
$p\in M^*$ and any $m\in k M$.
The fusion rules for such
fields are the addition of momenta (modulo $kM$),
$$A_p A_{\pr p}=A_{p+\pr p}.\e$$
Clearly the fusion ring is generated by $A_{m_i}$ where the $m_i$ are
the generators of the dual lattice $M^*$. The fusion algebra is the
group algebra over the abelian group $M^* \mod k M$. Now, a well
known theorem states that any abelian group is a direct sum of
cyclical groups, and so,
$${M^*\over k M}\approx Z_{n_1} \oplus Z_{n_2} \oplus \ldots \oplus
Z_{n_p},\e $$
where $Z_n$ stands for the order $n$ cyclical group and $n_i|n_{i+1}$.
Now, let $\lambda_i\in M^*$ be the generator of the $Z_{n_i}$ group. Then since
the fusion algebra is cyclical it follows that the relations in this algebra
are,
$$A_{\lambda_i}^{n_i}=1 \qquad {\rm for\ } i=1,2,\ldots,p.\e$$
Clearly, these relations can be integrated to a potential. Denoting by
$x_i=A_{\lambda_i}$, we find that the relations eq. (11) are the derivatives
with respect to $x_i$ of the potential,
$$V=\sum_{i=1}^p {x_i^{n_i+1}\over n_i+1}-x_i\e$$
The extrema of $V$ are the solutions of the relations $x_i^{n_i}=1$.
These can be parametrized by the elements $q\in M^* \mod kM$,
$$A_p=e^{2\pi i pq/k}.\e$$
which is precisely the matrix of modular transformations for the
bosonic model, in agreement with eq. (4).
We conclude that any bosonic model is described by a potential
which is given by eq. (12).
\par
Now the affine model at level $k$ may be considered as a folding of the
bosonic model, as is clear from eq. (6), where
one takes the lattice to be the long root lattice of the group, $M_l$,
at level $k+g$. Thus the problem of finding an affine fusion
potential reduces to finding a Weyl invariant bosonic potential.
Consider the symplectic algebra $C_n$. The simple roots of the
algebra are given by $\alpha_i={1\over\sqrt2}(\epsilon_i-
\epsilon_{i+1})$ for $i=1,2,\ldots,n-1$, and $\alpha_n=\sqrt2
\epsilon_n$, where $\epsilon_i$ are orthogonal unit vectors.
The long root lattice is spanned by $\sqrt2\epsilon_i$,
whereas the dual of it is spanned by $\epsilon_i/\sqrt2$.
It follows that $M^* \mod k M_l$ is the abelian group $Z_{2k}^n$.
In agreement with eq. (13) the fusion variety is given by
the points $q\in
M^*\mod kM_l$ according to
$A_p=e^{2\pi i pq}$. Denoting by $q=\sum r_i\epsilon_i/(\sqrt2 k)$, where
$r_i=0,1,2,\ldots,2k-1\mod2k$, we find that
$$q_i=A_{\epsilon_i/\sqrt2}=e^{2\pi i r_i/(2k)}\e$$
and there are $(2k)^n$ solutions.
\par
Let us look now for a potential which has precisely these points as
extrema. Let $q_i=\exp(\epsilon_i/\sqrt2)$. Consider the potential
$$V=\sum_i q_i^{n+k+1} +q_i^{-n-k-1},\e$$
which is the character of the fundamental representation to the
$(n+k+1)$th `power', in analogy with the $SU(N)$ case. Note that we
shifted $k$ to $n+k+1$ in view of the application to the affine case.
The extrema points
of $V$ are given by $\partial V/\partial q_i=0$, or
$$(k+n+1)q_i^{k+n}-(k+n+1)q_i^{-n-k-2}=0.\e$$
and coincide precisely with the points of $M^*\mod (k+n+1) M_l$, eq. (14).
Further, this potential has the essential property that it is invariant
under the action of the Weyl group, which is of the form
$\epsilon_i\rarrow \pm \epsilon_{p(i)}$ where $p$ is any permutation.
This is clear as $V$ is a scaled character of a representation.

Next, we need to express the potential $V$ in terms of Weyl invariant
variables, which are the generators of the affine fusion ring.
These we take, as mentioned earlier, to be the fundamental
representations expressed in terms of their characters,
$$x_\Lambda=\sum_{\lambda\in L(\Lambda)} e^{\lambda \vec\theta},\e$$
where $\Lambda$ is any fundamental weight, and $L(\Lambda)$ is the
set of weights in the representation, and $\vec \theta$ is some
specialization. An equivalent set of generators is,
$$y_r=2^r\sum_{i_1<i_2<\ldots <i_r} \cos\theta_{i_1} \cos\theta_{i_2}
\ldots\cos\theta_{i_r},\e$$
for $r=1,2,\ldots,n$.
The angular variables $\theta_i$ are defined by $2\cos\theta_i=
q_i+q_i^{-1}$. Note that $y_r$ is identical with the character
$x_r=\ch_{\Lambda^r}(\epsilon_i\theta_i\sqrt2)$, eq. (17), where $\Lambda_r$ is
fundamental weight,
up to terms which are of lower order in $q$, i.e., $y_r=x_r+
f(x_1,x_2,\ldots,x_{r-1})$ where $f$ is some function. It follows that
$x_r$ and $y_r$ are an equivalent set of generators, with a Jacobian
between them
which is equal to one. The Jacobian for the change of variables
$y_r \rarrow \theta_r$ is easy to compute using the Vandermonde
determinant,
$$J={\partial y_r\over \partial \theta_i}=2^{n(n+1)/2} \prod_{r=1}^n
\sin\theta_r \prod_{1\leq i<j\leq n} (\cos \theta_i-\cos\theta_j).\e$$
Note that the Jacobian is anti--invariant with respect to the Weyl
group, $\theta_i\rarrow\pm \theta_{p(i)}$. Thus it must be proportional
to the denominator of the Weyl character formula, which is the unique
generator of Weyl anti--invariant polynomials,
$$D=\sum_{w\in W} (-1)^w e^{w(\rho) \theta_i\epsilon_i\sqrt2},\e$$
Comparing the leading term shows that the two are in fact
identical, $D=J$.
It follows that the Jacobian $J$ vanishes at precisely the weights
$\lambda=\theta_i\epsilon_i\sqrt2$ which are fixed by
a Weyl reflection $\lambda=w(\lambda)$ for some reflection $w$.
Thus, these points are no longer extrema of the potential when
expressed in terms of the symmetric variables,
$$x_r={\sum_{w\in W} (-1)^w e^{2\pi i \omega(\Lambda_r+\rho)
\theta_i\epsilon_i\sqrt2}\over D}.\e$$
Further, points in the variety $M^*\mod (k+g) M_l$ which are related by
an element of the Weyl group clearly give the same value for $x_r$.
Thus the extrema of the potential
$$V=2\sum_{i=1}^n \cos (k+g) \theta_i,\e$$
when expressed in terms of the variables $x_r$ are given by the the Weyl
anti--invariant points in the variety $\theta_i=r_i/(2k+2g)$ where
$r_i=0,1,\ldots, 2(k+g)-1$. As is well known \REF\Kac{V. Kac,
``Infinite dimensional Lie algebras'', Cambridge, Cambridge University
Press (1985).}(e.g., ref. \r\Kac), these points are
in correspondence with the integrable highest weights at level $k$,
denoted by $\Lambda$,
$$\sum_i r_i\epsilon_i/\sqrt2={\Lambda+\rho\over k+g}.\e $$

To make the correspondence explicit, let us choose the representatives
$1\leq r_1 <r_2<\ldots <r_n\leq k+n$. It is clear that these are
representatives modulo $W$ of the extrema points. The fundamental weights
of $C_n$ are $\Lambda_r=(\epsilon_1+\epsilon_2+\ldots+\epsilon_r)/\sqrt2$,
and $\rho=\lb n \epsilon_1+(n-1)\epsilon_2+\ldots+\epsilon_n\rb/\sqrt2$.
It follows that
$$\sum_i r_i\epsilon_{n+1-i}/\sqrt2=\rho+\Lambda=\sum_i s_i \Lambda_i,\e$$
where $s_i=r_i-r_{i-1}-1$ label the fundamental weights. In particular,
$\sum s_i=r_n-n\leq k$, implying
that $\Lambda=s_i \Lambda_i$ is an integrable highest weight at level $k$.
Thus we find that the points of the variety on which the potential
$V$ has an extrema, are in correspondence with the integrable highest
weight fields at level $k$. Denoting a point by $l_\Lambda$,
where $\Lambda$ is an integrable highest weight, the value of the
primary field $\Mu$ on such a point is (from eq. (7))
$$x_{\Mu}=\ch_\Mu\left({\Lambda+\rho\over k+g}\right),\e$$
It
follows that these are precisely the correct values for the fusion
variety,
and so this ring is indeed the fusion ring of $C_n$.
\par
As a consequence of the fact that the Jacobian is the same as the
denominator, it follows that the representations in the theory form
a system of orthogonal polynomials when expressed in terms of the
generators, where the measure is equal (up to a constant) to the
denominator of the Weyl formula. This follows from
$$\twoline{C\delta_{\lambda,\mu}=\int_0^{2\pi}\ldots\int_0^{2\pi}
\prod_i d\phi_i \sum_{w,\pr w} (-1)^{w\pr w} e^{2\pi iw(\lambda+\rho)
\phi_i\epsilon_i\sqrt2}
e^{2\pi i\pr w(\mu+\rho)\phi\epsilon_i\sqrt2}=}
{\int_0^{2\pi}\ldots\int_0^{2\pi} \prod
\d{\phi_i} \ch_\lambda (\phi_i) \ch_\mu(\phi_i) D^2(\phi_i),}$$
where $C$ is some constant.
Changing variables to $x_r$ and remembering that the Jacobian is
$D(\phi_i)$ it follows
$$\int_C D(x_i)\prod_i \d{x_i} \ch_\lambda(x_i)\ch_\mu(x_i)=C\delta_{\lambda,
\mu}.\e$$
where $\lambda$ and $\mu$ are any highest weights.

Let us turn now to examples. In the case of $C_1\approx SU(2)$ the
potential we described for $C_n$, eq. (15), is immediately seen to be
identical to that of $SU(2)$ described earlier, eq. (3), which
is a Chebyshev polynomial of the first kind. The primary fields are
given by Chebyshev polynomials of the second kind, $P_m(2\cos \theta)=
2\sin(m+1)\theta/\sin \theta$ for $m=1,2,\ldots,k+1$. These indeed form an
orthogonal polynomial system with a measure which is equal to the
denominator, $D=\sqrt{4-x^2}$. Here $x$ stands for the fundamental
representation.

Consider now the general case. In order to compute the potential $V(y_i)$
it is convenient to use the following recursion relation (in analogy with
the $SU(N)$ case). The quantities $q_i=\exp(\theta_i)$ and $q_i^{-1}$ are
the solutions of the polynomial equation,
$$0=P=\prod_{i=1}^n (q-q_i)(q-q_i^{-1})=\sum_{r=0}^{2n} (-1)^rs_r
q^r.\e$$
The coefficients $s_r$ are Weyl--symmetric functions of the $q_i$ and can
be readily expressed in terms of the $x_i$, eq. (21). For example,
in the case of $n=2$ ($C_2$) we find
$$P=q^4-x_1 q^3+(x_2+1)q^2-x_1q+1=0.\e$$
Multiplying eq. (28) by $q^r$ for some $r$ and summing over the solutions,
we find the recursion relation for the potentials,
$$\sum_{i=0}^{2n} (-1)^i s_i V_{i+r}=0.\e$$
For $n=2$ this recursion relation becomes $V_{r+4}-x_1 V_{r+3}
+(1+x_2)V_{r+2}-x_1V_{r+1}+V_r=0$. Substituting the two initial
values $V_0=4$, $V_1=x_1$, using also $V_r=V_{-r}$, we find
recursively the potentials, $V_2=x_1^2-2x_2-2$, $V_3=x_1^3-3x_1x_2$,
$V_4=x_1^4-4x_1^2x_2+2x_2^2+4x_2-2$, $V_5=x_1^5-5x_1^3x_2+5x_1x_2^2
+5x_1x_2-5x_1$, etc.

According to the foregoing discussion the fusion ring of $C_n$ at level
$k$ is given by $P\lb x_i\rb/I_k$, where the ideal $I_k$ is generated by the
derivatives of the potential $I_k=(\partial_i V_{k+g}(x_i))$.
For $n=2$ ($g=3$) we find at $k=1$ the derivatives $\partial_{x_2}
V_4=-4x_1^2+4x_2+4=0$ and $\partial_{x_1}V_4=4x_1^3-8x_1x_2=0$.
Equivalently these relations are $x_1x_2=x_1$ and $x_1^2=x_2+1$. These
are precisely the relations of the fusion ring of $C_2\approx B_2$ at
level one, which are the same as the Ising model.
For a general $k$ the following relations can be established,
$$\eqalign{ {1\over k+3} {\partial V_{k+3}\over\partial x_2}&=
-\chi_{k+1,0}(x_1,x_2) \cr
{1\over k+3} {\partial V_{k+3}\over \partial x_1}&=x_1\chi_{k+1,0}(x_1,x_2)
-\chi_{k,1}(x_1,x_2)}\e$$
where $\chi_{n,m}(x_1,x_2)$ is the character of the representation
with highest weight $n\Lambda_1+m\Lambda_2$. It follows that the
ideal $(\partial_i V_{k+3})$ is generated by the characters
$\chi_{k+1,0}$ and $\chi_{k,1}$. These are precisely the two generators
of the relations of the fusion ring. Similarly, this type of relation
can be seen for $C_n$ with $n\geq3$.

The fusion potentials for $SU(N)$ as well as $C_n$ described above,
have an alternative description as superpotentials of massive $N=2$
supersymmetric Wess--Zumino theories (or Landau--Ginsburg) theories.
These are described by the lagrangian,
$${\cal L}=\int \d\theta\d{\bar\theta} K(x_i,x_i^*)+\int \d\theta
V(x_i)+{\rm c.c.}\e$$
where $x_i$ is interpreted as an $N=2$ scalar chiral superfield.
The chiral fields in the theory, again form an associative algebra,
the relations of which are described by the equation of motion
${\partial V\over\partial x_i}=0$. Thus we find that the operator
products of the chiral fields are described by the same ring
$P\lb x_i\rb/(\partial_i V)$. If we take $V$ to be a fusion potential,
the products of the chiral fields would be described by the fusion
coefficients. One can deform the $N=2$
supersymmetric field theory by adding the supersymmmetry partner of some
chiral field in the theory. This changes the superpotential $V$ by adding
to it the same chiral field, expressed as a polynomial in the $x_i$.
The $N=2$ LG theory is conformally
invariant if the potential $V$ admits some grading, i.e., a conserved
$U(1)$ charge with some particular assignment for the charges of the
generating chiral fields $x_i$ which appear in the lagrangian.
Thus the fusion rings may be considered as massive perturbations
of $N=2$ superconformal field theories. For example,
the potential for $SU(2)_4$ is given by the fourth Chebyshev polynomial
$T_4(x)=x^4-4x^2+2$. The $k=2$ minimal model has a superpotential which is
$V=x^4$. Thus the $SU(2)_4$ theory is a massive perturbation of
the $k=2$ minimal model. Similarly, the $SU(N)_k$ fusion rings may
be thought of as massive deformations \r{\FR,\CV}\ of the $N=2$ superconformal
theories based on the hermitian symmetric space construction (h.s.s.)
${SU(n+1)\times SO(2n)\over SU(n)\times U(1)}$ described in ref.
\REF\KS{Y. Kazama and H. Suzuki, Nucl. Phys. B321 (1989) 232.}\r\KS.

It turns out that the fusion rings of $C_n$ describe massive perturbations
of the same $SU(n)$ hermitian symmetric spaces, though with entirely
different perturbations. (Except, of course, in the case of
$C_1\approx SU(2)$.) To see this, assume that the variables $q_i$ in
eq. (14) all have a degree which is equal to one. The non--homogenous
terms in the potential for $q_i$ are due to the $q_i^{-1}$ terms in the
potential and in the variables $x_i$. Thus going to the conformal
limit amounts to dropping these terms, leaving us with the
potential
$$V=\sum_{i=1}^n  q_i^{n+k+1},\e$$
expressed in terms of the variables
$$x_r=\sum_{i_1<i_2<\ldots< i_r} q_{i_1} q_{i_2}\ldots q_{i_r}.\e$$
This is exactly the potential of the h.s.s. theory $SU(n+1)_k/SU(n)\times
U(1)$ \r\FR.
The degree of the $x_r$ variable is $r$. It is not hard to see
that the degrees of the perturbations in the theory are given by
$d_v-2r$ where $r$ is a positive integer and $d_v$ is the degree
of the potential. This is in contrast with the $SU(n+1)_k$ fusion
rings where the degree of the perturbations is $d_v-(n+1)r$. Thus,
these are two very different perturbations.

As an example consider $(C_2)_2$. The potential for this theory is
$V=x_1^5-5x_1^3x_2+5x_1x_2^2+5x_1x_2-5x_1$. The quasi--homogenous
part of this potential $V_0=x_1^5-5x_1^3x_2+5x_1x_2^2$ is
precisely the superpotential of the h.s.s. model $SU(3)/SU(2)\times U(1)$
at level $k=2$ (which is equivalent to the $k=10$ minimal model with
the $D$ modular invariant). The rest is the perturbation which is
$5x_1 x_2-5x_1$.

There are reasons to believe that fusion rings lead to integrable
$N=2$ supersymmetric models. The examples which were investigated
include the bosonic models ref. \REF\Fendley{P. Fendley W. Lerche, S.D. Mathur
and N.P. Warner, Nucl. Phys. B348 (1991) 66; W. Lerche and N.P. Warner,
Nucl. Phys. B358 (1991) 571.}\r\Fendley\ and the
$SU(N)_k$ case \REF\Nemesh{D. Nemeschansky and N.P. Warner,
Caltech preprint, USC--91--031, October (1991)}\r\Nemesh\
(Chebyshev perturbation). Thus the question arises whether this is
true for the general fusion field theory, and in particular for the
symplectic fusion rings described here. In order to examine this
question, we will compute the metric along the perturbation and show
that it is given as a solution of $\hat A_1^n$ Toda equation.

Let $\ket i$ denote the Ramond vacua of the theory.
The metric is described by the amplitude, $g_{\bar j,i}=\bra{\bar j} i\rangle
$.
Here $\bra{\bar j}$ denotes the vacuum obtained from the
$j$ state by spectral flow. Following ref. \r\CV\ let us denote by
$C_i$ the matrix representing the chiral field $C_i$, $(C_i)_{jk}=
N^i_{jk}$ where $N^i_{jk}$ are the structure constants of the chiral
algebra. For a one parameter potential $W=W(\lambda,x_i)$, the
infinitesimal perturbation is $W_\lambda={\partial W/\partial \lambda}$.
For such a perturbation, the metric obeys the following
differential equation
\REF\Strominger{A. Strominger, Com. Math. Phys. 133 (1990)
163}\r{\Strominger,\CV},
$$\bar\partial(g\partial g^{-1})+\lb W_\lambda,gW_\lambda^\dagger
   g^{-1} \rb=0,\e$$
where the equation is written in a matrix notation.
In addition, one imposes the reality condition which follows from
the transposition of chiral fields,
$$\eta^{-1}g(\eta^{-1}g)^*=1,\e$$
where $\eta$ is the transpose pairing,
$$\eta_{i,j}={\rm Res} {C_i C_j\over \prod_i \partial_i W}.\e$$
Consider the superpotential,
$$W=\lambda V(x_i) \e$$
where $V(x_i)$ is the potential for some fusion ring of a rational
conformal field theory. The known such fusion potentials, as described
above correspond to bosonic models and the $SU(N)$ and $C_n$
WZW models. As we shall shortly establish, in all such
fusion rings, the metric assumes a very particular form and
eq. (35) becomes an affine Toda equation for some root system.
As we shall argue, this in turn is an indication of the
integrability of the associated $N=2$ supersymmetric model, along
this line of perturbation.

Now, for the fusion potential $W=\lambda V(x)$ one can choose
as a basis for the chiral fields the actual primary fields of
the theory, i.e., the polynomials which represent the primary
fields in the original rational conformal field theory.
Moreover, one can use this basis for all
the values of the parameter $\lambda$. In this basis, rather
strikingly, the metric is diagonal, $g_{\lambda,\mu}=0$
unless $\lambda=\mu$. This can be argued, on the basis of the
fact that the metric is essentially a topological quantity and
the topological theory still preserves the charge conjugation
properties of the original rational conformal field theory.
This can also be seen by a direct calculation using eq. (35).
The technical reason for it is that $W$ acts as an external automorphism
on the primary fields.

Consider the fusion potential $V(\phi_i)$. It has extrema
at the points $\partial_i V=0$, which have the set of solutions
$x_1,x_2,\ldots,x_m$. The point basis for the chiral algebra
is defined by
$$l_i(x_j)=\delta_{i,j}\e$$
According to eq. (4) the primary fields are given in this basis by
$$\Phi_i=\sum_j S_{i,j}^\dagger/S_{0,i} l_j,\e$$
where $\Phi_i$ is the primary field and $S$ is the matrix of
modular transformations. In the point basis, the potential
$V$ acts diagonally and has the value $V(x) l_i=V(x_i)l_i$.
In the primary field basis it gives
$$V(x) \Phi_i=\sum_j S_{i,j}^\dagger/S_{0,j} V(x_j) l_j.\e$$
Consider, for example, the $SU(N)$ potential eq. (3). Substituting
the values of the variety we find that $V(x_\lambda)=\exp(
2\pi i\lambda\Lambda_1)$ where $\lambda$ is the integrable weight
which correspond to the point, and $\Lambda_1$ is the first
fundamental weight, or $V(x_\lambda)$ is given by the central
element which correspond to $\lambda$. Using the relation
$$S_{\lambda,\sigma(\mu)}^\dagger=e^{2\pi i\Lambda_\sigma\lambda}
S_{\lambda,\mu}^\dagger,\e$$
where $\sigma$ is an external automorphism and $\Lambda_\sigma$
is the corresponding fundamental weight, we find that eq. (41)
assumes the form,
$$V \Phi_\mu=\Phi_{\sigma(\mu)},\e$$
where $\sigma$ is the generator of external automorphisms.
Similarly, for a gaussian model, $V=x^{n+1}/(n+1)-x$, we find an
analogous expression where $\sigma$ is replaced by the fusion
automorphism, $\sigma(m)=m+1 \mod n$. This particular form of
the potential is responsible for the simplicity of the equation
for the metric. Indeed we find that eq. (35) takes the form
$$\bar \partial(g_\mu\partial g_\mu^{-1})+ g_\mu^{-1}
g_{\sigma(\mu)}-g_{\sigma^{-1}(\mu)}^{-1}g_\mu=0.\e$$
Thus the metric is diagonal, and its flow involves only elements related
by an external automorphism. Since $\sigma$ is an element of order $n$,
generating a $Z_n$ cyclical group, it follows that eq. (44) is precisely
the $\hat A_n$ Toda theory, as is seen by the change of variables
$g_{\sigma^i(
\mu)}=e^{\phi_i}$, for $i=0,1,\ldots,n$,
$$\partial\bar\partial \phi_i=e^{\phi_{i+1}-\phi_i}-e^{\phi_i-\phi_
{i-1}},\e$$
which is an equivalent form of the $\hat A_n$ Toda theory. This was
already established in \r\CV\ in a slightly different form.

Consider now the $C_n$ fusion potential, eq. (15).
The extrema points are given by eq. (16), $q_i=\exp[\pi ir_i/(k+n+1)]$,
where $r_i=0,1,\ldots,2(k+n)+1$ modulo $k+n+1$. Thus, it follows that the
values of the potential at these points are given by,
$$V(r_1,r_2,\ldots,r_n)=\sum_{i=1}^n (-1)^{r_i}.\e$$
Now, unlike the previous cases, $V$ does not correspond to a central
element. However, it still corresponds to an automorphism of the
fusion rules. Remembering that the choice of representatives which
corresponds to the integrable highest weights is $1<r_1<r_2<\ldots<r_n$,
the automorphism assumes the form $r_i\rarrow k+n+1-r_i$. Using the
diagonal ansatz $g_{\lambda,\mu}=g_\lambda\delta_{\lambda,\mu}$ (which
can be justified by an explicit calculation), along with the assumption
that the metric factorizes into a product of the metrics for the
separate $SU(2)$ factors that make this potential. ($C_n$ appears here
as a symmetrized product of the $n$ $SU(2)$'s, as follows from eq.
(33)). For a connection with Chern-Simons--Witten theory
see the appendix. Thus the equation for the metric assumes the
form
$$\twoline{\bar\partial(g_{r_1,r_2,\ldots,r_n} \partial g_{r_1,r_2,\ldots,r_n}
^{-1})+\sum_{i=1}^n g_{r_1,r_2,\ldots,k+n+1-r_i,\ldots,r_n}
g^{-1}_{r_1,r_2,\ldots,r_n}-}{g_{r_1,r_2,\ldots,r_n}g_{r_1,r_2,\ldots,
k+n+1-r_i,\ldots,r_n}^{-1}=0.}$$
Consistency of the ansatz implies that the solution can be written as
$$g_{r_1,r_2,\ldots,r_n}= {1\over(\lambda\bar\lambda)^{n/2}}
\prod_{i=1}^n e^{L(r_i)},\e$$
where $L(r_i)$ is a solution of the sinh-gordon equation,
$$\partial\bar\partial L(r_i)=2\sinh [2L(r_i)].\e$$
The transpose pairing $\eta$ is computed from eq. (37) and is found
to be
$$\eta_{\sigma,\mu}={1\over\lambda^n}
\delta_{\sigma,\bar\mu},\e$$
where $\bar\mu$ is defined as sending $r_i$ to $k+n+1-r_i$,
for all $i$.
The reality condition eq. (36) implies $L(r_i)=-L(k+n+1-r_i)$.
The solutions of the sinh--gordon equation are classified by their
boundary condition at the origin,
$$L=\half t\log z+\half s+O(\lambda^{2-|t|}{\bar\lambda}^{2-|t|})\e$$
where $t$ and $s$ are some real constants, and $z=2\sqrt{\lambda\bar\lambda}$.
The condition for the
solution to have no poles on the real axis ($\lambda\bar\lambda>0$)
is
$$e^{s/2}={1\over 2^{t/2}} {\Gamma(\half-{t\over4})\over\Gamma(\half
+{t\over4})}.\e$$
Since the solution is expected to have no singularities on the real
axis, it remains only to fix the the value of $t(r)$ for $r=1,2,\ldots,
k+n$ to specify it uniquely. This we do by examining
the behaviour for $\lambda\rarrow0$. With the change of variables
$x_r\rarrow \lambda^{r/(k+n+1)} x_r$ the potential eq. (15) becomes that
of the superconformal theory $SU(n+1)/SU(n)$ as discussed earlier, up to
a perturbation which vanishes with $\lambda$. Under such a change of
variables the primary field $P_{r_1,r_2,\ldots,r_n}$ becomes
the field $\lb a_1,a_2,\ldots,a_k \rb$ in the h.s.s. theory where
$$[a_i]=[1,1,\ldots,1,2,2,\ldots,2,\ldots,n,n,\ldots,n],\e$$
and where $i$ appears $r_i-r_{i-1}-1$ times.
When expressed in terms of the $x_i$ this field becomes the polynomial
(Giambelli's formula)
$$\lb a_1,a_2,\ldots,a_k\rb=\det_{1\leq i<j\leq k} x_{a_i+i-j}.\e$$
This can be seen by directly examining the recursion relation for
the polynomials $P_{r_1,r_2,\ldots,r_n}$ which is
$$y_r P_{r_1,r_2,\ldots,r_n}=\sum_{i_1<i_2<\ldots<i_r\atop\pm} P_{r_1,
r_2,\ldots,r_{i_1}\pm1,\ldots,r_{i_2}\pm1,\ldots,r_n},\e$$
where the $\pm$ are summed over independently.
This is a special case of the general recursion relation,
$$\ch_\lambda \ch_\mu=\sum_{\rho\in L(\lambda)} \ch_{\mu+\rho},\e$$
which holds for any group and any weights $\lambda$ and $\mu$
provided that $\mu$ is sufficiently large (i.e., having $\mu+\rho$ as
a dominant weight for all $\rho$). Eq. (56) can be proved by
writing explicitly the formula for the characters, using the Weyl character
formula.
The limit $\lambda\rarrow0$ corresponds precisely to
taking the homogenous part of the Pieri--like formula eq. (55),
which then becomes precisely the classical Pieri formula,
implying, in turn, the Giambelli formula eq. (54). This shows
that indeed, at this limit, the $C_n$ polynomials are given by the
determinant eq. (54).

{}From the fact that the two point function of this field is finite
at the limit $\lambda\rarrow0$ it follows that
$$g_{r_1,r_2,\ldots,r_n}=(\lambda\bar\lambda)^{c/3-q/(k+n+1)}
\langle [a_1,a_2,\ldots,a_k]| [a_1,a_2,\ldots,a_k]\rangle,
\e$$
where $c$ is the central charge of the theory, $c/3=\sum_i 1/2-i/(k+n+1)$,
and $q/(k+n+1)$ is the $U(1)$ charge of the field, $q=\sum  a_i$.
Thus we find for the boundary condition of the solution $L(r)$,
$${t(r)\over4}=\half-{r\over k+n+1},\e$$
and so the normalization  becomes (using eq. 51--52),
$$\langle [a_1,a_2,\ldots,a_k] | [a_1,a_2,\ldots
,a_k] \rangle=\prod_{i=1}^n {\Gamma(r_i/(k+n+1))\over\Gamma(1-
r_i/(k+n+1))}.\e$$
This result for the normalizations of the h.s.s. superconformal theory
agrees in the case of $n=1$ with those of the minimal models, the only ones
known so far.
In addition, the normalizations are explicitly dual, i.e.,
they remain invariant with the exchange of $n$ and $k$, provided one
replaces the $r_i$'s with the ones corresponding to the transpose of the
original Young tableau. This is precisely the duality of the h.s.s. theory
$SU(n+1)_k/SU(n)\times U(1)$.

Let us consider now the opposite limit where $\lambda\bar\lambda\rarrow
\infty$. In this limit the sinh--gordon equation behaves asymptotically
as
$$L(r)=-\sqrt{2/\pi}(\lambda\bar\lambda)^{-1/4}\sin[\pi t(r)/4]
e^{-4\sqrt{\lambda\bar\lambda}}.\e$$
Now, in the point basis $l_i$, eq. (39), the metric assumes the form
$$G_{\lambda,\mu}=S_{0,\lambda}^\dagger S_{0,\mu} \sum_{\rho}
S_{\lambda,\rho}^\dagger S_{\mu,\rho} g_{\rho,\rho},\e$$
where $G_{\lambda,\mu}=\langle l_\lambda| l_\mu\rangle$ is the
point basis metric and $S$ is the matrix of modular transformations.
$S$ may be written as
$$S_{r_i;m_i}=\sum_{p\in S_n} (-1)^p \prod_{i=1}^n
\sin\left({r_i m_{p(i)}\over n+k+1} \right),\e$$
up to an overall constant determined by unitarity,
where $p\in S_n$ is any permutation and $(-1)^p$ is $\pm1$
if $p$ is even or odd. Substituting the asymptotic form of $L$
and using the matrix $S$ we find the asymptotics of the metric
in the point basis,
$$g_{r_i;m_i} \propto \sum_p (-1)^p \sum_{\gamma_i=\pm1,0}
e^{-4\sqrt{\lambda\bar\lambda}\sum\gamma_i^2}
\delta_{m_{p(i)}-r_i-\gamma_i}.\e$$
The asymptotics of the metric has a universal form \r\CV
$$g_{i,j}={\beta_{i,j}\over (\lambda\bar\lambda)^{n/2}} e^{-2M_{i,j}},\e$$
where $M_{i,j}=|W(x_i)-W(x_j)|$ is the mass of the soliton interpolating
between the two classical vacua described by the point basis \r\Nemesh,
and $\beta_{i,j}$ is the transition amplitude between the vacua.
We thus infer that in the $C_n$ massive theory there are $n$ different
soliton
masses which are $M_s=2s$, for $s=1,2,\ldots,n$.
Eq. (63) further describes the allowed
transitions which are $r_i\rarrow r_i\pm 1,0$ up to a permutation of the
$r_i$, where there are exactly $s$ transitions for the $s$ soliton.
Note that even though naively many other transitions could
be allowed, resulting in other solitons, the calculation of the metric
shows that these are the only ones present\foot{Note that the fact that
$\beta_{ij}$ vanishes implies only that $g_{ij}$ is smaller
asymptotically, through the subleading terms. These, however, correspond
to multi--soliton transitions, and can be ignored in reading the
solitonic structure.}.
\par
The kink structure that arises in the $N=2$ fusion theories
described above affords the following common description. The set
of ground states, being identical to the fusion variety described earlier,
corresponds to the integrable highest weights at
level $k$.
The kinks that interpolate between these ground states, correspond
to the tensor product with one of the fundamental weights, i.e.,
the weights $\lambda$ and $\mu$ are connected by a kink if and only
if the fusion coefficient $N_{\lambda,\Lambda_i}^\mu$ is
nonzero for some fundamental weight $\Lambda_i$. The different $\Lambda_i$
correspond to different kink masses. In the case of $SU(2)$ the calculation
of the metric \r\CV\ shows indeed such a kink structure. For $SU(N)$
it is presumed that the solution of the affine Toda equation, eq. (45),
does so as well, though a precise analysis is hindered by the lack of
treatment of the boundary conditions. Finally, for
$C_n$ the kink structure we calculated is identical with the
multiplication by the different fundamental representations of $C_n$.
This can be seen by noting that the soliton transition amplitude,
eq. (63), is identical to the Giambelli formula, eq. (55)
where the representation label, $r$ is identified with the $s$ soliton.
Further, each representation corresponds to a distinct soliton mass, which
is $2s$.
Thus it is natural to assume that
for all the groups the kink structure of the fusion theory assumes
this form. Each kink is described by some fundamental representation, and
the allowed transitions are given by the fusion rules with that representation.

Further, knowing the kink structure of the theory, along with the
fact that it is integrable, allows one to write down a consistent
set of factorizable $S$ matrices which correspond to it, by
solving the constraints of the Yang--Baxter equation, crossing symmetry
and unitarity for the kink scattering theory. Amazingly, precisely the same
kink structure described above was considered in ref.\REF\DVF{H.J. de
Vega and V.A. Fateev, Int. Journal of Mod. Phys. A6 (1991) 3221}
\r\DVF\ where it was suggested to describe the perturbed coset models
$G_k\times G_1/G_{k+1}$ by the field $\Phi_{{\rm ad},0}$,
in the case of $G=SU(N)$. In addition the masses of the kinks are the same
as described here, being given by $|W(x_i)-W(x_j)|$.
Since the analysis of ref. \r\DVF\ depends only on the kink structure it
follows
that the $S$ matrices described there apply also for the fusion conformal
field theories. At first this may seem puzzling since these are
perturbations of different conformal field theories (i.e., the supersymmetric
h.s.s. models) with a different central charge. The explanation of this is that
while the minimal $S$ matrices indeed describe the perturbed $G_k\times G_1/
G_{k+1}$ models, the solution allows additional factors, the so called $Z$
factors, that when appropriately chosen give the full $S$ matrix of the
fusion theories. In fact, in the case of $SU(N)_1$ the perturbation becomes
gaussian, and can be used to demonstrate this. As was discussed in ref.
\r\Fendley\ the $S$ matrices are those of affine
$\hat A_n$ Toda theories with some imaginary coupling, which are a
non--minimal version of the
$A_n$ $S$ matrices. The minimal $S$ matrices of \r\DVF\ reduce
for $k=1$ to these $A_n$ $S$ matrices. Thus, it is necessary only to
determine the precise
$Z$ factors for the $S$ matrices of ref. \r\DVF\ to obtain the full
$S$ matrices. This argumentation applies equally well for all other
groups, where it is expected that the kinks will similarly correspond
to the fundamental representations fusing between the classical vacua,
which are given by the integrable weights at level $k$.
The $S$ matrices
still should be given by a quantum group construction, similar
to the case of $SU(N)$.
This will be
discussed elsewhere\foot{Further support for this picture is found in
ref. \REF\FI{P. Fendley and K. Intriligator, BUHEP91, HUTP-91/A043,
preprint, November (91)}\r\FI\ where the case of $SU(2)$ is treated
in detail.
Our suggested $S$ matrices agree with those of \r\FI\ for $SU(2)$.}.
Note the triple role played here by the fusion rules. First they
are used to define the potential and chiral algebra of the theory.
In addition, the metric turns out to be diagonal in the primary
field basis, for all the couplings. Finally, the kink structure of
the theory and its associated $S$ matrix turns out to be given
by the very same fusion rules!

The fact that the metric assumes this particular simple form
is a considerable evidence for the fact that this fusion field
theory is integrable. Note, in particular, the close analogy
between the $C_n$ case and the $C_1=SU(2)$ case. The integrability of
the latter has been investigated in ref. \r\Nemesh\
at the level of finding conserved currents. It
is expected that, similarly, conserved currents could be found for
the $C_n$ theories described here.

This work on the symplectic fusion rings gives further support to two
conjectures regarding the structure of fusion rings \r\FR\ and its connection
to integrable $N=2$ supersymmetric theory \r\CV. Indeed, we find that
the fusion rings of $C_n$ are described by a potential for any level,
in accordance with the conjecture of ref. \r\FR. Further, these potentials
lead to Toda equations for the metric, suggesting the
integrability of the associated $N=2$ supersymmetric field theory.

Note added: After the completion of this paper we became aware of two
related works. In ref. \REF\Riggs{M. Bourdeau, E.J. Mlawer, H. Riggs and H.J.
Schnitzer, Brandeis University preprint, BRX-TH-327, October (1991).}\r\Riggs\
the fusion potentials
of $C_n$ are also described. In ref. \REF\Nem{D. Nemeschansky and N.P. Warner,
to appear.}\r\Nem\ the lattice $S$ matrices associated to fusion rings
are investigated. Our results appear to be in agreement with both works.

\ack
It is a pleasure for us to acknowledge discussions with S. Elitzur,
D. Kazhdan, A.B. Zamolodchikov and Al.B. Zamolodchikov. The work of
A. Schwimmer was supported in part by BSF grant no. 89--00140.
D. Gepner wishes to thank the Rockefeller University and especially
M. Evans and N. Khuri for the hospitality during the final
stages of this work.

\def \sr{\sqrt{2}}
\def\r#1{$[{\rm #1}]$}
\Appendix{A.}
\noindent{\bf The connection with CSW theory.}

The fusion rings have a natural setting in the framework of
Chern--Simons--Witten (CSW) theories
\REF\jones{E. Witten, Comm. Math. Phys. 121 (1989) 351.}\r\jones.
Quantising the CSW theory for a group $G$ at level $k$ on a torus,
one obtains a finite dimensional Hilbert space. This space has a very
convenient description in terms of an effective quantum mechanical
problem \REF\shmuel{S. Elitzur, G. Moore, A. Schwimmer and N. Seiberg,
Nucl. Phys. B326 (1989) 108.}\r\shmuel, in which the conjugate variables are
$a_1$ and $a_2$,
$$[a_1^i,a_2^j]={2\pi i \over k+g}\delta^{ij}. \e$$
The indices $i,j$ take the  values $1,2,...,r$ where $r$ is the rank of the
group. The wave functions of the states are
$$\Psi^{eff}_\mu    (\vec a_1)
=\sum_{w\in W}(-1)^w\delta^P(\vec a_1-
    {w(\mu    +\rho) \over k+g} ), \e$$
where
$$\delta^P(\vec a_1-\vec \ell)\equiv
\sum_{\alpha\in  M_l     }\delta(\vec a_1-\vec\ell-    \alpha),\e$$
and where $\mu $ is a weight and $M_l$ is the long root lattice.
The independent
states are in one to one correspondence with the integrable highest
weight representations of the affine Lie algebra at level $k$.
 The natural `gauge invariant' operators in the reduced phase space
(i.e., invariant under shifts in $M_l$ and under the action of the Weyl group)
are the Weyl characters
 $$ x_ \lambda \equiv  ch_ \lambda (a_1)={\sum_{w\in W} (-1)^w
  e^{2\pi i \omega(\lambda  +\rho) a_1} \over
 \sum_{w\in W} (-1)^w e^{2\pi i \omega(\rho)a_1}},\e$$
where $\lambda$ is a fundamental weight. Similar operators can be defined
in terms of $a_2$.
The $x_ \lambda$ are related to the Wilson loops on the
torus in the full CSW theory.
Obviously, the set of $x_{\lambda}$, for $\lambda$ which is a fundamental
weight,
generate a commutative ring generated by the $x_\lambda$ operators,
which are expressed in terms
of the commuting $a_1$. However, since the latter act on a finite dimensional
Hilbert space, there are relations among the generators of this ring
which fulfill `secular equations'. These equations are of the general form,
$$P(x_{\lambda_1},\ldots,x_{\lambda_r})\Psi_\mu =0,\e$$
for all the $\mu $ which are compatible with eq. (A.2).
Therefore, the ideal is defined by a set of polynomials of $x_\lambda$
  which have
as solutions precisely all the $x_{\lambda_i}^\mu$ defined by
$$x_{\lambda_i}^\mu=ch_{\lambda_i}({\mu+\rho\over k+g}).\e$$

 \def\sr{{\sqrt2}}
  The SU(2) potential is,
 $$V(x_1)={1\over2}ch_1((k+2)a_1)=2\cos(2\pi{(k+2)a_1\over \sr}),\e$$
where $k$ is the level.
 The corresponding polynomial equation is
 $${\sin(2\pi{(k+2)a_1\over\sr})\over \sin( 2\pi {a_1\over \sr})}
      =0.\e$$
 The action of the potential eq. (A.7) on an arbitrary representation
 $$ch_m(a_1)\equiv {\sin(2\pi{(m+1) a_1\over \sr})\over
   \sin(2\pi{a_1\over \sr})},\e$$
 where $ m=0,1,2,\ldots,k$, is
 $$\twoline{V(x_1) ch_m(a_1)=\cos(2\pi{(k+2)a_1\over \sr}) \ch_m(a_1)-}{
   \cos(2\pi{(m+1) a_1\over \sr}) \ch_{k+1}(a_1)=-\ch_{k-m}(a_1),}$$
 where we used eq. (A.8) in order to add the second term.
  For SU(N) the potential is given by eq. (3). The polynomial
 equations give rise to the conditions
$$q^{N+k}_1=q^{N+k}_2=..=q^{N+k}_N \;\;{\rm and} \;\;
      q^{N(N+k)}_1=1.\e$$
 Using eq. (A.11) and the definition of the character eq. (A.4),
 the action of $V$ on any character can be easily calculated,
$$V(\vec a_1 )\ch_{\vec \mu} (\vec a_1 )=
    N\ch_{\vec \mu +(N+k) \vec \lambda _1}(\vec a_1 ),\e$$
 where $\vec\lambda _1$ is the fundamental weight corresponding
 to the basic representation. The automorphism eq. (A.12) is cyclic
 of order $N$ due to the second condition in (A.11).

For the symplectic groups $C_n$, using a decomposition of $\vec a_1$ in
an orthogonal basis $\epsilon_1$, one finds a close connection with
$SU(2)^n$. The potential is given by
$$V(\vec a_1)=\sum_{i=1}^n cos(2\pi{(k+n+1)a_1^i \over\sr})\e$$
 and therefore the polynomial equations are
$$\ch_{k+n}(a_1^i)=0 \;\;{\rm
  for} \;i=1,...n.\e$$
However, due to the presence of the Jacobian between the $C_n$
 and the $SU(2)^n$ characters, eq. (A.14)
 acts on a Hilbert space completely {\it antisymmetrised}
in terms of the $a_1^i$.
  We label a general irreducible representation of $C_n$ by the
integer components $r_i$ in the ${\epsilon_i\over \sr}$ basis
of the highest weight $\lambda$. Then the character of the representation
 becomes
$$\ch_{\vec\lambda} (\vec a_1)={\sum_{P} (-1)^P  \prod_{j=1}^n
 \ch_{r_j+n-j}(a_1^{j_l}) \over\sum_{P} (-1)^P\prod_{i=1}^n \ch_{n-i}
 (a_1^{i_m})},\e$$
  where $P$ denotes the permutations of the $a_1^i$.
     Applying repeatedly eq. (A.12), we find the action of $V$ on
the representations
$$V_{\vec\lambda} (\vec a_1)=
 -\sum_{i=1}^n \ch_{\vec \lambda^{(i)}}
 (\vec a_1),\e$$
 where
 the components of $\vec \lambda^{(i)}$ are  $r_1,..,k+2i-1-n-r_i,..
  ,r_n$, where $r_1,r_2,\ldots,r_n$ are the components of $\vec\lambda$.
    In order to scale the variables $x_s \equiv \ch_{\vec \lambda_s}$,
 where $\vec \lambda_s$ is the fundamental weight corresponding to
  $r_1=r_2=...=r_s=1$, all the others$=0$, as $\lambda^{r\over{k+n+1}}$,
    we have to scale the individual $SU(2)$ variables $\ch_1(a_1^j)$
  as $\lambda^{1\over (k+n+1)}$. This follows from the explicit
  expression of the character eq. (A.15). Therefore, in the conformal limit
 we obtain a $SU(2)^n$ theory defined, however, in the CSW
 formalism, completely antisymmetrised in the $a_1^i$ Hilbert space.
  Once this restriction        is imposed, one can use factorisation
  in $SU(2)$ to simplify the calculation of the $C_n$ metric, as is done
in the main text.
\refout

\bye